\documentclass[aps,pre,graphicx,reprint,superscriptaddress]{revtex4-1}

\usepackage{graphicx,amsmath,amsthm,amssymb,bm,subfloat}

\draft 

\begin{document}


\title{Formation of current singularity in a topologically constrained plasma} 


\author{Yao Zhou}
\email[]{yaozhou@princeton.edu}
\affiliation{Plasma Physics Laboratory and Department of Astrophysical Sciences, Princeton University, Princeton, New Jersey 08543, USA}
\author{Yi-Min Huang}
\affiliation{Plasma Physics Laboratory and Department of Astrophysical Sciences, Princeton University, Princeton, New Jersey 08543, USA}
\author{Hong Qin}
\affiliation{Plasma Physics Laboratory and Department of Astrophysical Sciences, Princeton University, Princeton, New Jersey 08543, USA}
\affiliation{Department of Modern Physics, University of Science and Technology of China, Hefei, Anhui 230026, China}

\author{A.\,Bhattacharjee}
\affiliation{Plasma Physics Laboratory and Department of Astrophysical Sciences, Princeton University, Princeton, New Jersey 08543, USA}

\date{\today}

\begin{abstract}
Recently a variational integrator for ideal magnetohydrodynamics in Lagrangian labeling has been developed. Its built-in frozen-in equation makes it optimal for studying current sheet formation. We use this scheme to study the Hahm-Kulsrud-Taylor problem, which considers the response of a 2D plasma magnetized by a sheared field under sinusoidal boundary forcing. We obtain an equilibrium solution that preserves the magnetic topology of the initial field exactly, with a fluid mapping that is non-differentiable. Unlike previous studies that examine the current density output, we identify a singular current sheet from the fluid mapping. These results are benchmarked with a constrained Grad-Shafranov solver. The same signature of current singularity can be found in other cases with more complex magnetic topologies.

\end{abstract}

\pacs{}

\maketitle 


\section{Introduction.}
Current sheet formation has long been an issue of interest in plasma physics.
In toroidal fusion plasmas, closed field lines exist at rational surfaces. It is believed that current singularities are inevitable when these surfaces are subject to resonant perturbations \cite{Grad1967,Rosenbluth1973,Hahm1985,Cary1985,Hegna1989,Bhattacharjee1995,Boozer2010,Dewar2013,Helander2014,Loizu2015}, which jeopardizes the existence of 3D equilibria with nested flux surfaces. In the solar corona, field lines are tied into the boundaries and do not close on themselves. Yet Parker \cite{Parker1972,Parker} argued that there would still be current sheets forming frequently, and the subsequent field line reconnections can lead to substantial heating. This theory has stayed controversial to this day \cite{VanB,Zweibel,Longcope1994,Ng1998,Bogoyavlenskij2000,Low2006,Huang2009,Janse2010,Low2010,Longbottom1998,Craig2005,Pontin2005,Craig,Craig2014}.

Albeit inherently a dynamical problem, current sheet formation is usually treated by examining magnetostatic equilibria for simplicity. The justification is, if the final equilibrium that an initially smooth magnetic field relaxes to contains current sheets, they must have formed during the relaxation. Here the plasma is supposed to be perfectly-conducting, so the equilibrium needs to preserve the magnetic topology of the initial field. This topological constraint is difficult to explicitly describe and attach to the magnetostatic equilibrium equation, and to enforce it is a major challenge for studying current sheet formation, either analytically or numerically. 

It turns out this difficulty can be overcome by adopting Lagrangian labeling, where the frozen-in equation is built into the equilibrium equation, instead of the commonly used Eulerian labeling. Analytically this was first shown by Zweibel and Li (ZL) \cite{Zweibel}. Numerically, a Lagrangian relaxation scheme has been developed using conventional finite difference \cite{Craig1986}, and extensively used to study current sheet formation  \cite{Longbottom1998,Craig2005,Pontin2005,Craig,Craig2014,Craig2005b}. 
It has later been found that its current density output can violate charge conservation ($\nabla\cdot\mathbf{j}=0$), and mimetic discretization has been applied to fix it \cite{Pontin2009,Candelaresi2014}. 

Recently, a variational integrator for ideal magnetohydrodynamics (MHD) in Lagrangian labeling \cite{Zhou2014} has been developed using discrete exterior calculus \cite{Desbrun}. It is derived in a geometric and field-theoretic manner such that many of the conservation laws of ideal MHD, including charge conservation, are naturally inherited. Here we present the first results of applying this novel scheme to studying current sheet formation.

We consider a problem first proposed by Taylor and studied by Hahm and Kulsrud (the HKT problem from here on), where a 2D plasma in a sheared magnetic field is subject to sinusoidal boundary forcing \cite{Hahm1985}. It was originally designed to study forced magnetic reconnection induced by resonant perturbation on a rational surface. In the context of studying current sheet formation, we refer to finding a topologically constrained equilibrium solution to it as the ideal HKT problem. ZL's linear solution to this problem \cite{Zweibel} contains a current sheet but also a discontinuous displacement that is unphysical. It has remained unclear whether the nonlinear solution to it is ultimately singular or smooth.

We study how the nonlinear numerical solution to the ideal HKT problem converges with increasing spatial resolution, and find the fluid mapping along the neutral line non-differentiable. Unlike previous studies that depend heavily on the current density diagnostic that is more vulnerable to numerical inaccuracies \cite{Longbottom1998,Craig2005,Pontin2005,Craig,Craig2014,Craig2005b}, we identify a singular current sheet from the quadratic fluid mapping normal to the neutral line.
Prompted by these results, we employ a Grad-Shafranov solver where the equilibrium guide field is not prescribed \textit{a priori} but constrained by flux conservation \cite{Huang2009} to independently verify the accuracy of our Lagrangian method.

\section{The HKT problem.} 
The HKT problem \cite{Hahm1985} originally considers a 2D incompressible plasma magnetized by an equilibrium field $B_y = \epsilon x$ with constant shear $\epsilon$. The boundaries at $x=\pm a$ are then subject to sinusoidal perturbations so that $x=\pm (a-\delta\cos ky)$. One branch of the perturbed equilibrium solutions with no magnetic islands along the neutral line $x=0$ reads
\begin{equation}
B_y = \epsilon [x +\text{sgn}(x)ka\delta\cosh kx\cos ky/\sinh ka].\label{HK}
\end{equation}
Note that the sign function $\text{sgn}(x)$ introduces a jump in $B_y$, namely a current sheet at the neutral line. However, it can be shown that this solution introduces residual islands with width of $O(\delta)$ on both sides of the neutral line \cite{Boozer2010,Dewar2013}. Its magnetic topology is therefore different from that of the initial field, which makes it not a topologically constrained equilibrium. 

This is not surprising since solution (\ref{HK}) is obtained by solving the magnetostatic equilibrium equation,
\begin{equation}
(\nabla\times\mathbf{B})\times\mathbf{B}=\nabla p,\label{JcrossB}
\end{equation}
where $p$ is the pressure. A given set of boundary conditions usually allows for more than one solutions to this equation, and additional information is needed to identify a specific one. Often it is prescribed to the equilibrium, such as the pressure and guide field profiles in the Grad-Shafranov equation \cite{Grad1967}. Yet the information distinguishing a topologically constrained solution from others is the very constraint to preserve the initial magnetic topology, which is mathematically challenging to explicitly attach to Eq.\,(\ref{JcrossB}) and its solutions \cite{Janse2010,Low2010}.

However, this topological constraint can be naturally enforced if one adopts Lagrangian labeling, which traces the motion of the fluid elements in terms of a continuous mapping from the initial position $\mathbf{x}_0$ to the current position $\mathbf{x}(\mathbf{x}_0,t)$. In this formulation, the advection (continuity, adiabatic, and frozen-in) equations are \cite{Newcomb}
\begin{subequations}\label{advection}
\begin{align}
\rho\,\mathrm{d}^3{x}=\rho_0\,\mathrm{d}^3x_0&\Rightarrow \rho=\rho_0/J,\label{continuity}\\
p/\rho^\gamma=p_0/\rho_0^\gamma&\Rightarrow p=p_0/J^\gamma,\label{adiabatic}\\
B_i\,\mathrm{d}S_i=B_{0i}\,\mathrm{d}S_{0i}&\Rightarrow B_i=x_{ij}B_{0j}/J,\label{frozenin}
\end{align}
\end{subequations}
where $x_{ij}=\partial x_i/\partial x_{0j}$, $J=\det(x_{ij})$ is the Jacobian, $\gamma$ the adiabatic index, $\rho_0=\rho(\mathbf{x}_0,0)$ the initial mass density, and the same goes for $p_0$ and $\mathbf{B}_0$.
They reflect the fact that in ideal MHD, mass, entropy, and magnetic flux are advected by the motion of the fluid elements. They are built into the ideal MHD Lagrangian and the subsequent Euler-Lagrange equation \cite{Newcomb},
\begin{align}
&\rho_0\ddot{x}_i-B_{0j}\frac{\partial}{\partial x_{0j}}\left(\frac{x_{ik}B_{0k}}{J}\right)\nonumber\\
&+\frac{\partial J}{\partial x_{ij}}\frac{\partial }{\partial x_{0j}}\left(\frac{p_0}{J^\gamma}+\frac{x_{kl}x_{km}B_{0l}B_{0m}}{2J^2}\right)=0.\label{momentum3}
\end{align}
This is the momentum equation, the only ideal MHD equation in Lagrangian labeling. 

Without time dependence, Eq.\,(\ref{momentum3}) becomes an equilibrium equation. Its solutions will satisfy not only Eq.\,(\ref{JcrossB}) but automatically the topological constraint in studying current sheet formation, since the initial field configuration $\mathbf{B}_0$ is built in.
In contrast, not all solutions to Eq.\,(\ref{JcrossB}) can necessarily be mapped from given initial conditions. Thus, the equilibrium equation in Lagrangian labeling offers a more natural and mathematically explicit description for the problem of current sheet formation, which simply becomes whether there exist singular solutions to such equation, given smooth initial and boundary conditions. If the initial field $\mathbf{B}_0$ is smooth, any singularity in the equilibrium field $\mathbf{B}$ should trace back to that in the fluid mapping $\mathbf{x}(\mathbf{x}_0)$. 

ZL first used the advantageous Lagrangian labeling to study current sheet formation \cite{Zweibel}. Their linear solution to the ideal HKT problem reads 
\begin{align}\label{linear}
\left(\xi_x,\xi_y\right) = \left\{f(x_0)\cos ky_0,-\left[f'(x_0)/k\right]\sin ky_0 \right\},
\end{align}
where $f=-\text{sgn}(x_0)a\delta\sinh kx_0/(x_0\sinh ka)$ and $\bm{\xi}$ is the displacement.
It agrees with Eq.\,(\ref{HK}) linearly and contains also a current sheet at the neutral line, but discontinuity in $\xi_x$ as well, which is not physically permissible. The failure at the neutral line is expected from the linear solution since the linear assumption breaks down there. Similar discontinuity in the displacement also appears in a linear analysis of the internal kink instability \cite{Rosenbluth1973}.

It is worth noting that instead of enforcing incompressibility ($J=1$), ZL used a guide field $B_{0z}=\sqrt{1-\epsilon^2x_0^2}$ so that the unperturbed equilibrium is force-free. Their solution (\ref{linear}) turns out to be linearly incompressible ($\nabla\cdot\bm{\xi}=0$). Even near the neutral line, the plasma should still be rather incompressible because the guide field dominates there. Therefore, the physics of the ideal HKT problem is not affected by such alteration in setup, which we shall adopt in our numerical studies. 

\section{Numerical results.} 
The numerical scheme we use is a recently developed variational integrator for ideal MHD \cite{Zhou2014}. It is obtained by discretizing Newcomb's Lagrangian for ideal MHD in Lagrangian labeling \cite{Newcomb} on a moving mesh. Using discrete exterior calculus \cite{Desbrun}, the momentum equation (\ref{momentum3}) is spatially discretized into a conservative many-body form $M_i\ddot{\mathbf{x}}_i=-\partial V/\partial \mathbf{x}_i$, where $M_i$ and $\mathbf{x}_i$ are the mass and position of the $i$th vertex, respectively, and $V$ is a spatially discretized potential energy. When the system is integrated in time, friction may be introduced to dynamically relax it to an equilibrium with minimal $V$. The scheme inherits built-in advection equations from the continuous formulation, and thus avoids the error and dissipation associated with solving them. It has been shown that the scheme can handle prescribed singular current sheets without suffering from artificial field line reconnection. Such capability of preserving the magnetic topology makes it an optimal tool for studying current sheet formation. 

\begin{figure}
\includegraphics[scale=0.46]{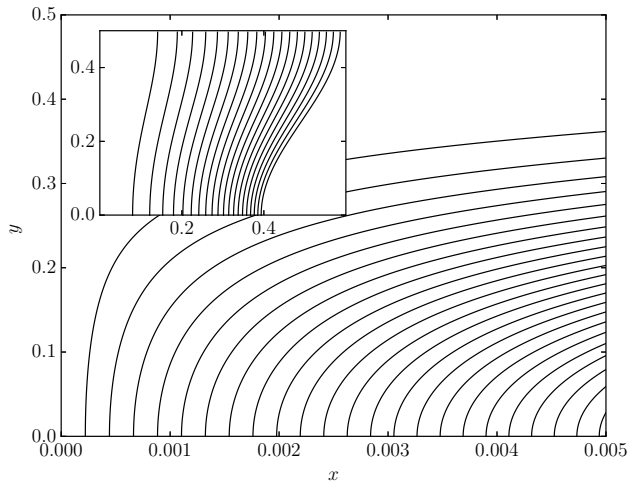}
\caption{\label{magnetic} Equilibrium field line configuration in the vicinity of the neutral line, and the entire domain (inset). The field lines appear equally spaced along $y=0$ near the neutral line.}
\end{figure}

For the ideal HKT problem, we use a structured triangular mesh. Thanks to the symmetry in this problem, we can simulate only a quarter of the domain, $[0,a]\times[0,\pi/k]$. At $x_0=a$ it is constrained that $x=a-\delta\cos ky$. The vertices are allowed to move tangentially along but not normally to the boundaries. These boundary conditions are exactly consistent with the original HKT setup. The parameters we choose are $\epsilon=1$, $\rho_0=1$, $a=0.5$, $k=2\pi$, and $\delta=0.1$. We use a large perturbation so that the nonlinear effect is more significant and easier to resolve. The vertices are distributed uniformly in $y$ but non-uniformly in $x$ in order to devote more resolution to the region near the neutral line. The system starts from a smoothly perturbed configuration consistent with the boundary conditions and relaxes to equilibrium. In Fig.\,\ref{magnetic} we plot the field line configuration of the equilibrium.

An observation from Fig.\,\ref{magnetic} is that $B_y(x,0)$ becomes a finite constant near the neutral line. To better illustrate the origin of such tangential discontinuity, we review a simple yet instructive 1D problem \cite{Craig2005b} where an exact nonlinear solution with current sheet is available. Consider the same sheared field $B_{0y}=\epsilon x_0$ as in the HKT problem, but the plasma is compressible, with no guide field or pressure present. The boundaries at $x_0=\pm a$ are perfectly conducting walls. The system is not in equilibrium and will collapse toward a topologically constrained one with a quadratic fluid mapping $x=x_0|x_0|/a$. The Jacobian $J=2|x_0|/a$ is zero at the neutral line, where the equilibrium field $B_y=B_{0y}/J=\epsilon a\,\text{sgn}(x)/2$ yields a current sheet. As we shall show next, the current sheet in the ideal HKT problem develops from the same ingredients, sheared initial field and quadratic fluid mapping.

\begin{figure}
\includegraphics[scale=0.46]{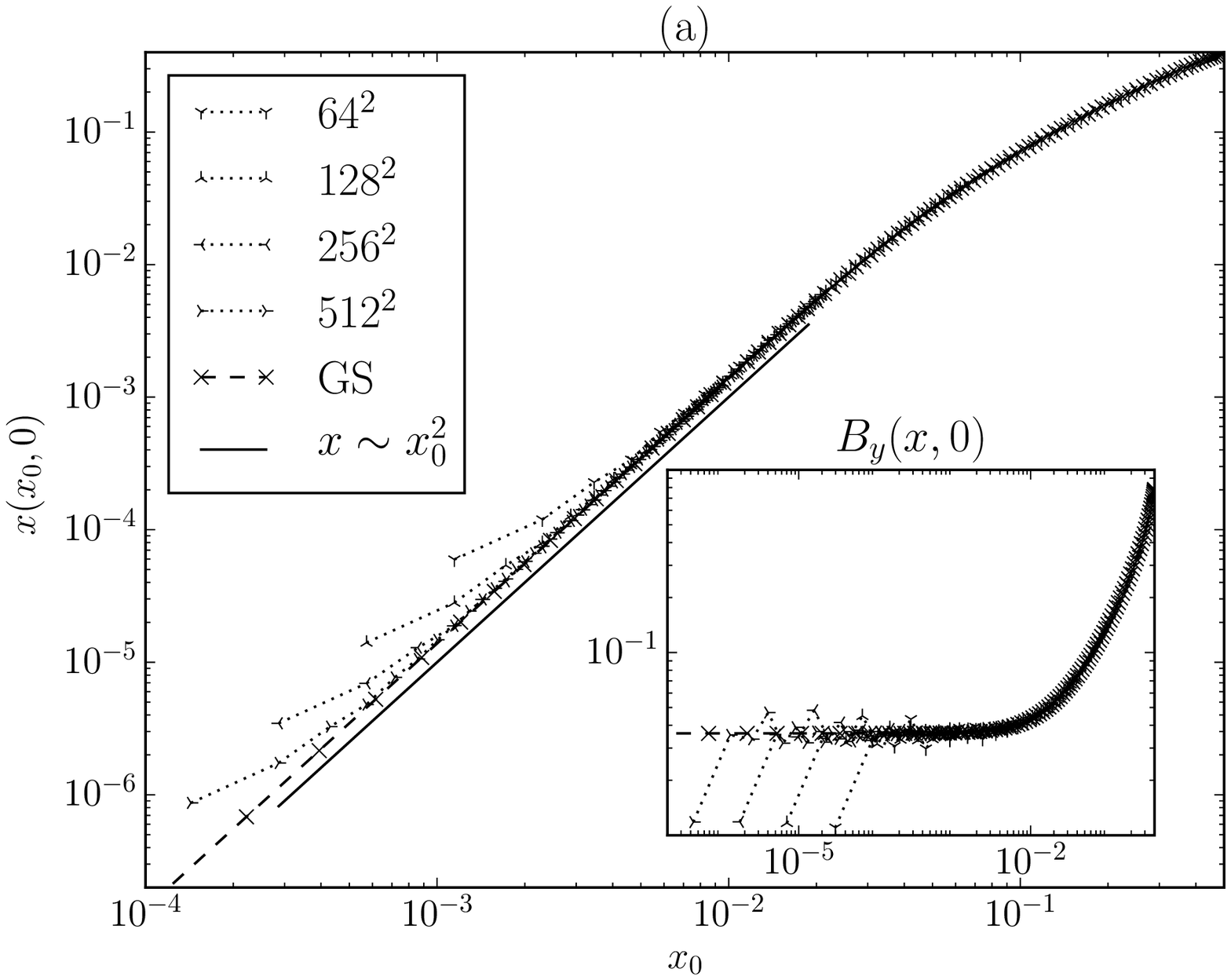}
\includegraphics[scale=0.46]{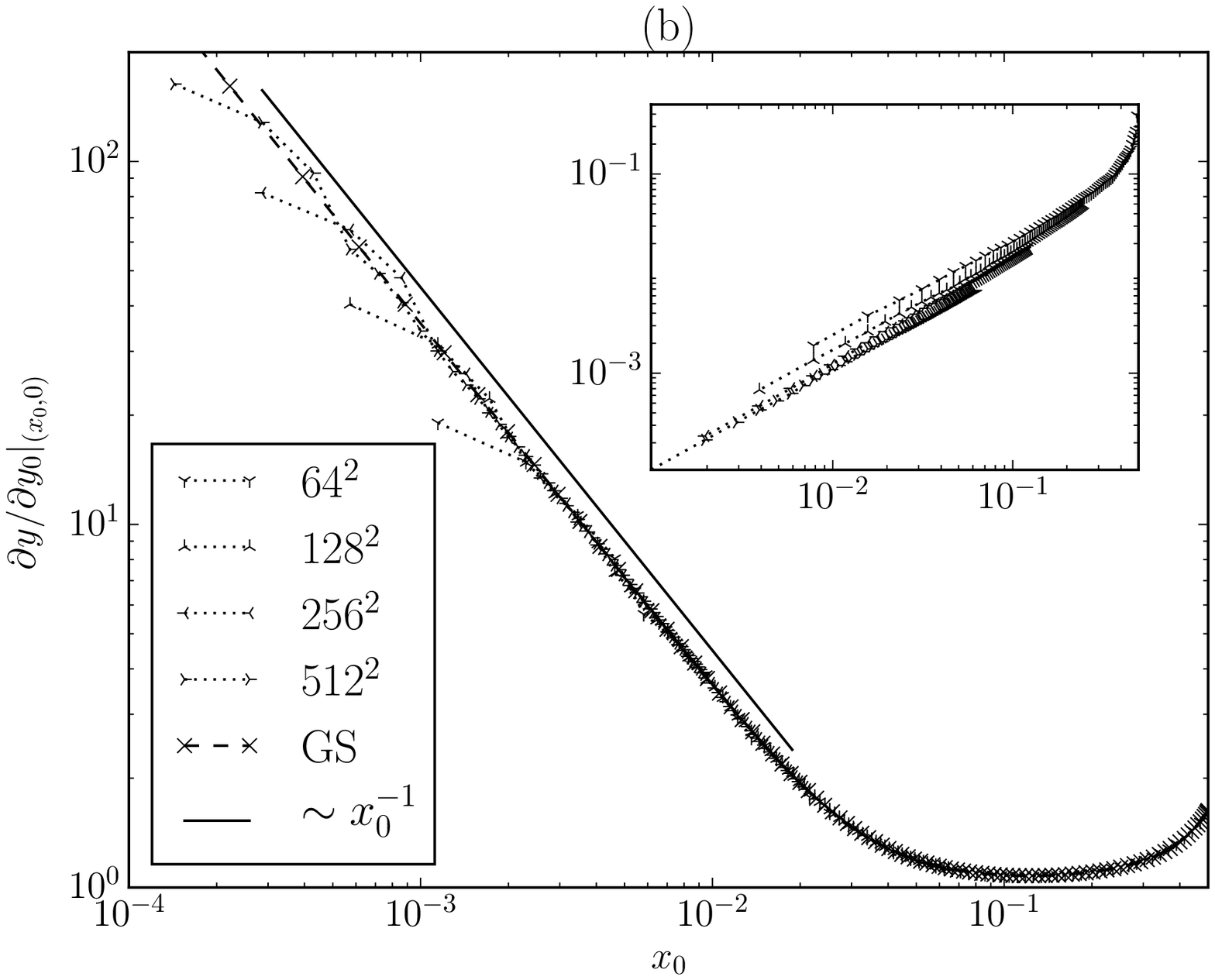}
\caption{\label{mapping} Numerical solutions of $x(x_0,0)$ (a), $B_y(x,0)$ (inset of a) and $\partial y/\partial y_0|_{(x_0,0)}$ (b) for different resolutions (dotted lines). The converged parts agree with the results obtained with a constrained Grad-Shafranov solver (dashed lines). Near the neutral line, $x(x_0,0)$ and $\partial y/\partial y_0|_{(x_0,0)}$ show $x_0^{2}$ and $x_0^{-1}$ power laws respectively, while $B_y(x,0)$ approaches a finite constant. The solutions do not converge for the few vertices closest to the neutral line. In the inset of (b), the final versus initial distance to $(0,0.5)$ for the vertices on the neutral line, i.\,e.\, $0.5-y(0,y_0)$ vs.\,$0.5-y_0$ for different resolutions are shown to not converge. Solutions with higher resolutions are shown in part.}
\end{figure}

We check how the equilibrium solutions converge with increasing spatial resolutions, from $64^2$ to $128^2$, $256^2$, and $512^2$. For solutions with higher resolutions, we only show the part in the vicinity of the neutral line, since they converge very well away from it. In Fig.\,\ref{mapping}(a), we plot the equilibrium fluid mapping normal to the neutral line at $y_0=0$, namely $x(x_0,0)$. For the part of the converged solutions near $x_0=0$, quadratic power law $x\sim  x_0^2$ can be observed. As discussed in the 1D case above, together with a sheared field $B_{0y}\sim x_0$, such a mapping leads to a magnetic field $B_y=B_{0y}/(\partial x/\partial x_0)\sim \text{sgn}(x)$ [note that $J=(\partial x/\partial x_0)(\partial y/\partial y_0)$ at ${y_0=0}$] that is discontinuous at $x_0=0$, as plotted in the inset of Fig.\,\ref{mapping}(a).

Despite the remarkable resemblance on the mechanism of current sheet formation, there is a key distinction between the 1D collapse and the ideal HKT problem. For the former, the plasma is infinitely compressible at the neutral line, and the equilibrium fluid mapping is continuous and differentiable. If there is guide field or pressure, no matter how small, to supply finite compressibility that prevents the Jacobian from reaching zero, the topologically constrained equilibrium would be smooth with no current sheet \cite{Craig2005b}. In the ideal HKT problem, the plasma is (close to) incompressible. This is confirmed by our numerical solutions which show $J\approx1+O(\delta^2)$. As a result, the equilibrium fluid mapping turns out to be non-differentiable.

At $y_0=0$, the converged power law $x\sim  x_0^2$ suggests that $\partial x/\partial x_0\sim  x_0$ would vanish as $x_0$ approaches $0$. To ensure incompressibility, there should be $\partial y/\partial y_0\sim  x_0^{-1}$ which would diverge at $x_0=0$. This is shown in Fig.\,\ref{mapping}(b). Physically, this means the fluid elements on the neutral line are infinitely compressed in the normal direction ($x$), while infinitely stretched in the tangential direction ($y$).

However, it is difficult to numerically resolve a diverging $x_0^{-1}$ power law at $x_0=0$. Therefore, the numerical solutions $x(x_0,0)$ and $\partial y/\partial y_0|_{(x_0,0)}$ both deviate from the converged power law for the few vertices closest to the neutral line. This deviation reduces but does not disappear with increasing resolutions. The inset of Fig.\,\ref{mapping}(b) shows that the vertices on the neutral line become more packed at $(0,0.5)$ as the resolution increases, suggesting that the solutions do not converge on the neutral line.

These numerical results are benchmarked with the solutions from a constrained Grad-Shafranov (GS) solver. In this solver the equilibrium guide field is determined self-consistently with a constraint to preserve its flux at each flux surface \cite{Huang2009}, unlike conventional ones where it is prescribed as a flux function. Without this feature the solver would not be capable for studying the ideal HKT problem. As shown in Fig.\,\ref{mapping}, the GS results are in excellent agreement with the converged part of those obtained with the Lagrangian scheme. Since the fluid mapping is inferred rather than directly solved for, the GS solver is able to achieve better agreement with the $x_0^{-1}$ power law shown in Fig.\,\ref{mapping}(b). However, it should be pointed out that the applicability of the GS solver is limited to 2D problems with nested flux surfaces, while the Lagrangian scheme can be readily generalized to problems with complex magnetic topologies or to 3D. In fact, with the Lagrangian scheme, we identify the same signature of current singularity as shown in Fig.\,\ref{mapping} when studying the coalescence instability of magnetic islands \cite{Finn1977,Longcope}, a problem the GS solver is not applicable to.

\section{Discussion.}
A straightforward conclusion we can draw from the numerical solutions to the ideal HKT problem is that there exists no smooth equilibrium fluid mapping. Nonetheless, this does not necessarily leads to the conclusion that there is a genuine current singularity. In the context of studying current sheet formation, one needs to take the further step of differentiating the mapping and confirm the existence of a possible current sheet. This is exactly what we have done in this paper. 

In previous studies that use similar Lagrangian relaxation
methods \cite{Longbottom1998,Craig2005,Pontin2005,Craig,Craig2014,Craig2005b}, current singularities are identified by examining whether the peak current density diverges with increasing spatial resolutions. However, involving second-derivatives, the output of current density is generally less reliable than that of the fluid mapping, especially at a current sheet where the mesh can be highly distorted. Since any singularity in the current density should trace back to that in the more fundamental fluid mapping, we choose to identify current singularities by examining the latter. In this paper, the current sheet we find originates from the quadratic fluid mapping normal to the neutral line. In this sense, we consider our numerical evidence for current sheet formation in 2D to be the strongest in the extant literature.

It is also worthwhile to compare our result with the recent work of Loizu \textit{et al}., which also studies the ideal HKT problem \cite{Loizu2015}, but in the context of finding well-defined ideal MHD equilibria with nested flux surfaces. For the original HKT setup, they find no such equilibrium. Then they introduce an alternate formulation to the problem, which in our terminology is equivalent to making the initial magnetic field discontinuous with $B_{0y}=\epsilon[x_0+\text{sgn}(x_0)\alpha]$, where $\alpha$ is a positive constant. Analytically, this would make the linear solution (\ref{linear}) continuous, such that smooth equilibrium fluid mapping becomes possible. We are able to get converged numerical solutions as well when such formulation is adopted. However, the results in this paper differ from those of Ref.\,\cite{Loizu2015} in that we begin with a smooth initial condition, rather than one with discontinuity, so as to observe the emergence of a current sheet. 

ZL \cite{Zweibel} studied the ideal HKT problem as a variation of Parker's original model where a uniform field in 3D line-tied configuration is subject to footpoint motions \cite{Parker1972}. Since a sheared field can be realized from a uniform field by sheared footpoint motion, it is more closely related to Parker's model than other variations that involve more complex field topology, such as magnetic nulls \cite{Craig,Craig2014,Pontin2005}. The dynamics also stay simple since there are no violent instabilities like the coalescence instability \cite{Longcope1994}. Now that we have confirmed that there is a current singularity in the 2D problem, naturally our next step is to find out whether it survives in 3D line-tied configuration. In fact, in Ref.\,\cite{Zweibel} it is conjectured that current sheets would not form in the 3D ideal HKT problem. 

\begin{acknowledgements}
We acknowledge helpful discussions with S.\,Hudson, C.\,Liu, J.\,Loizu, J.\,Squire, J.\,Stone, and E.\,Zweibel. This research was supported by the U.S. Department of Energy under Contract No.\,DE-AC02-09CH11466.
\end{acknowledgements}


%
%

%


\bibliography{HK.bib}

\begin{thebibliography}{35}%
\makeatletter
\providecommand \@ifxundefined [1]{%
 \@ifx{#1\undefined}
}%
\providecommand \@ifnum [1]{%
 \ifnum #1\expandafter \@firstoftwo
 \else \expandafter \@secondoftwo
 \fi
}%
\providecommand \@ifx [1]{%
 \ifx #1\expandafter \@firstoftwo
 \else \expandafter \@secondoftwo
 \fi
}%
\providecommand \natexlab [1]{#1}%
\providecommand \enquote  [1]{``#1''}%
\providecommand \bibnamefont  [1]{#1}%
\providecommand \bibfnamefont [1]{#1}%
\providecommand \citenamefont [1]{#1}%
\providecommand \href@noop [0]{\@secondoftwo}%
\providecommand \href [0]{\begingroup \@sanitize@url \@href}%
\providecommand \@href[1]{\@@startlink{#1}\@@href}%
\providecommand \@@href[1]{\endgroup#1\@@endlink}%
\providecommand \@sanitize@url [0]{\catcode `\\12\catcode `\$12\catcode
  `\&12\catcode `\#12\catcode `\^12\catcode `\_12\catcode `\%12\relax}%
\providecommand \@@startlink[1]{}%
\providecommand \@@endlink[0]{}%
\providecommand \url  [0]{\begingroup\@sanitize@url \@url }%
\providecommand \@url [1]{\endgroup\@href {#1}{\urlprefix }}%
\providecommand \urlprefix  [0]{URL }%
\providecommand \Eprint [0]{\href }%
\providecommand \doibase [0]{http://dx.doi.org/}%
\providecommand \selectlanguage [0]{\@gobble}%
\providecommand \bibinfo  [0]{\@secondoftwo}%
\providecommand \bibfield  [0]{\@secondoftwo}%
\providecommand \translation [1]{[#1]}%
\providecommand \BibitemOpen [0]{}%
\providecommand \bibitemStop [0]{}%
\providecommand \bibitemNoStop [0]{.\EOS\space}%
\providecommand \EOS [0]{\spacefactor3000\relax}%
\providecommand \BibitemShut  [1]{\csname bibitem#1\endcsname}%
\let\auto@bib@innerbib\@empty
\bibitem [{\citenamefont {Grad}(1967)}]{Grad1967}%
  \BibitemOpen
  \bibfield  {author} {\bibinfo {author} {\bibfnamefont {H.}~\bibnamefont
  {Grad}},\ }\href {\doibase 10.1063/1.1761965} {\bibfield  {journal} {\bibinfo
   {journal} {Physics of Fluids}\ }\textbf {\bibinfo {volume} {10}},\ \bibinfo
  {pages} {137} (\bibinfo {year} {1967})}\BibitemShut {NoStop}%
\bibitem [{\citenamefont {Rosenbluth}\ \emph {et~al.}(1973)\citenamefont
  {Rosenbluth}, \citenamefont {Dagazian},\ and\ \citenamefont
  {Rutherford}}]{Rosenbluth1973}%
  \BibitemOpen
  \bibfield  {author} {\bibinfo {author} {\bibfnamefont {M.~N.}\ \bibnamefont
  {Rosenbluth}}, \bibinfo {author} {\bibfnamefont {R.~Y.}\ \bibnamefont
  {Dagazian}}, \ and\ \bibinfo {author} {\bibfnamefont {P.~H.}\ \bibnamefont
  {Rutherford}},\ }\href {\doibase doi:10.1063/1.1694231} {\bibfield  {journal}
  {\bibinfo  {journal} {Physics of Fluids}\ }\textbf {\bibinfo {volume} {16}},\
  \bibinfo {pages} {1894} (\bibinfo {year} {1973})}\BibitemShut {NoStop}%
\bibitem [{\citenamefont {Hahm}\ and\ \citenamefont
  {Kulsrud}(1985)}]{Hahm1985}%
  \BibitemOpen
  \bibfield  {author} {\bibinfo {author} {\bibfnamefont {T.~S.}\ \bibnamefont
  {Hahm}}\ and\ \bibinfo {author} {\bibfnamefont {R.~M.}\ \bibnamefont
  {Kulsrud}},\ }\href {\doibase 10.1063/1.865247} {\bibfield  {journal}
  {\bibinfo  {journal} {Physics of Fluids}\ }\textbf {\bibinfo {volume} {28}},\
  \bibinfo {pages} {2412} (\bibinfo {year} {1985})}\BibitemShut {NoStop}%
\bibitem [{\citenamefont {Cary}\ and\ \citenamefont
  {Kotschenreuther}(1985)}]{Cary1985}%
  \BibitemOpen
  \bibfield  {author} {\bibinfo {author} {\bibfnamefont {J.~R.}\ \bibnamefont
  {Cary}}\ and\ \bibinfo {author} {\bibfnamefont {M.}~\bibnamefont
  {Kotschenreuther}},\ }\href {\doibase 10.1063/1.864973} {\bibfield  {journal}
  {\bibinfo  {journal} {Phys.Fluids}\ }\textbf {\bibinfo {volume} {28}},\
  \bibinfo {pages} {1392} (\bibinfo {year} {1985})}\BibitemShut {NoStop}%
\bibitem [{\citenamefont {Hegna}\ and\ \citenamefont
  {Bhattacharjee}(1989)}]{Hegna1989}%
  \BibitemOpen
  \bibfield  {author} {\bibinfo {author} {\bibfnamefont {C.~C.}\ \bibnamefont
  {Hegna}}\ and\ \bibinfo {author} {\bibfnamefont {A.}~\bibnamefont
  {Bhattacharjee}},\ }\href {\doibase 10.1063/1.859152} {\bibfield  {journal}
  {\bibinfo  {journal} {Physics of Fluids B: Plasma Physics}\ }\textbf
  {\bibinfo {volume} {1}},\ \bibinfo {pages} {392} (\bibinfo {year}
  {1989})}\BibitemShut {NoStop}%
\bibitem [{\citenamefont {Bhattacharjee}\ \emph {et~al.}(1995)\citenamefont
  {Bhattacharjee}, \citenamefont {Hayashi}, \citenamefont {Hegna},
  \citenamefont {Nakajima},\ and\ \citenamefont {Sato}}]{Bhattacharjee1995}%
  \BibitemOpen
  \bibfield  {author} {\bibinfo {author} {\bibfnamefont {A.}~\bibnamefont
  {Bhattacharjee}}, \bibinfo {author} {\bibfnamefont {T.}~\bibnamefont
  {Hayashi}}, \bibinfo {author} {\bibfnamefont {C.~C.}\ \bibnamefont {Hegna}},
  \bibinfo {author} {\bibfnamefont {N.}~\bibnamefont {Nakajima}}, \ and\
  \bibinfo {author} {\bibfnamefont {T.}~\bibnamefont {Sato}},\ }\href {\doibase
  10.1063/1.871369} {\bibfield  {journal} {\bibinfo  {journal} {Physics of
  Plasmas}\ }\textbf {\bibinfo {volume} {2}},\ \bibinfo {pages} {883} (\bibinfo
  {year} {1995})}\BibitemShut {NoStop}%
\bibitem [{\citenamefont {Boozer}\ and\ \citenamefont
  {Pomphrey}(2010)}]{Boozer2010}%
  \BibitemOpen
  \bibfield  {author} {\bibinfo {author} {\bibfnamefont {A.~H.}\ \bibnamefont
  {Boozer}}\ and\ \bibinfo {author} {\bibfnamefont {N.}~\bibnamefont
  {Pomphrey}},\ }\href {\doibase doi:10.1063/1.3507307} {\bibfield  {journal}
  {\bibinfo  {journal} {Physics of Plasmas}\ }\textbf {\bibinfo {volume}
  {17}},\ \bibinfo {pages} {110707} (\bibinfo {year} {2010})}\BibitemShut
  {NoStop}%
\bibitem [{\citenamefont {Dewar}\ \emph {et~al.}(2013)\citenamefont {Dewar},
  \citenamefont {Bhattacharjee}, \citenamefont {Kulsrud},\ and\ \citenamefont
  {Wright}}]{Dewar2013}%
  \BibitemOpen
  \bibfield  {author} {\bibinfo {author} {\bibfnamefont {R.~L.}\ \bibnamefont
  {Dewar}}, \bibinfo {author} {\bibfnamefont {A.}~\bibnamefont
  {Bhattacharjee}}, \bibinfo {author} {\bibfnamefont {R.~M.}\ \bibnamefont
  {Kulsrud}}, \ and\ \bibinfo {author} {\bibfnamefont {A.~M.}\ \bibnamefont
  {Wright}},\ }\href {\doibase doi:10.1063/1.4817276} {\bibfield  {journal}
  {\bibinfo  {journal} {Physics of Plasmas}\ }\textbf {\bibinfo {volume}
  {20}},\ \bibinfo {pages} {082103} (\bibinfo {year} {2013})}\BibitemShut
  {NoStop}%
\bibitem [{\citenamefont {Helander}(2014)}]{Helander2014}%
  \BibitemOpen
  \bibfield  {author} {\bibinfo {author} {\bibfnamefont {P.}~\bibnamefont
  {Helander}},\ }\href {\doibase 10.1088/0034-4885/77/8/087001} {\bibfield
  {journal} {\bibinfo  {journal} {Reports on Progress in Physics}\ }\textbf
  {\bibinfo {volume} {77}},\ \bibinfo {pages} {087001} (\bibinfo {year}
  {2014})}\BibitemShut {NoStop}%
\bibitem [{\citenamefont {Loizu}\ \emph {et~al.}(2015)\citenamefont {Loizu},
  \citenamefont {Hudson}, \citenamefont {Bhattacharjee},\ and\ \citenamefont
  {Helander}}]{Loizu2015}%
  \BibitemOpen
  \bibfield  {author} {\bibinfo {author} {\bibfnamefont {J.}~\bibnamefont
  {Loizu}}, \bibinfo {author} {\bibfnamefont {S.}~\bibnamefont {Hudson}},
  \bibinfo {author} {\bibfnamefont {A.}~\bibnamefont {Bhattacharjee}}, \ and\
  \bibinfo {author} {\bibfnamefont {P.}~\bibnamefont {Helander}},\ }\href
  {\doibase 10.1063/1.4906888} {\bibfield  {journal} {\bibinfo  {journal}
  {Physics of Plasmas}\ }\textbf {\bibinfo {volume} {22}},\ \bibinfo {pages}
  {022501} (\bibinfo {year} {2015})}\BibitemShut {NoStop}%
\bibitem [{\citenamefont {Parker}(1972)}]{Parker1972}%
  \BibitemOpen
  \bibfield  {author} {\bibinfo {author} {\bibfnamefont {E.}~\bibnamefont
  {Parker}},\ }\href@noop {} {\bibfield  {journal} {\bibinfo  {journal} {The
  Astrophysical Journal}\ }\textbf {\bibinfo {volume} {174}},\ \bibinfo {pages}
  {499} (\bibinfo {year} {1972})}\BibitemShut {NoStop}%
\bibitem [{\citenamefont {Parker}(1994)}]{Parker}%
  \BibitemOpen
  \bibfield  {author} {\bibinfo {author} {\bibfnamefont {E.~N.}\ \bibnamefont
  {Parker}},\ }\href@noop {} {\bibfield  {journal} {\bibinfo  {journal}
  {Spontaneous current sheets in magnetic fields: with applications to stellar
  x-rays. International Series in Astronomy and Astrophysics, Vol. 1. New York:
  Oxford University Press, 1994.}\ }\textbf {\bibinfo {volume} {1}} (\bibinfo
  {year} {1994})}\BibitemShut {NoStop}%
\bibitem [{\citenamefont {Van~Ballegooijen}(1985)}]{VanB}%
  \BibitemOpen
  \bibfield  {author} {\bibinfo {author} {\bibfnamefont {A.}~\bibnamefont
  {Van~Ballegooijen}},\ }\href@noop {} {\bibfield  {journal} {\bibinfo
  {journal} {The Astrophysical Journal}\ }\textbf {\bibinfo {volume} {298}},\
  \bibinfo {pages} {421} (\bibinfo {year} {1985})}\BibitemShut {NoStop}%
\bibitem [{\citenamefont {Zweibel}\ and\ \citenamefont {Li}(1987)}]{Zweibel}%
  \BibitemOpen
  \bibfield  {author} {\bibinfo {author} {\bibfnamefont {E.~G.}\ \bibnamefont
  {Zweibel}}\ and\ \bibinfo {author} {\bibfnamefont {H.-S.}\ \bibnamefont
  {Li}},\ }\href@noop {} {\bibfield  {journal} {\bibinfo  {journal} {The
  Astrophysical Journal}\ }\textbf {\bibinfo {volume} {312}},\ \bibinfo {pages}
  {423} (\bibinfo {year} {1987})}\BibitemShut {NoStop}%
\bibitem [{\citenamefont {Longcope}\ and\ \citenamefont
  {Strauss}(1994)}]{Longcope1994}%
  \BibitemOpen
  \bibfield  {author} {\bibinfo {author} {\bibfnamefont {D.~W.}\ \bibnamefont
  {Longcope}}\ and\ \bibinfo {author} {\bibfnamefont {H.~R.}\ \bibnamefont
  {Strauss}},\ }\href {\doibase 10.1086/175045} {\bibfield  {journal} {\bibinfo
   {journal} {The Astrophysical Journal}\ }\textbf {\bibinfo {volume} {437}},\
  \bibinfo {pages} {851} (\bibinfo {year} {1994})}\BibitemShut {NoStop}%
\bibitem [{\citenamefont {Ng}\ and\ \citenamefont
  {Bhattacharjee}(1998)}]{Ng1998}%
  \BibitemOpen
  \bibfield  {author} {\bibinfo {author} {\bibfnamefont {C.~S.}\ \bibnamefont
  {Ng}}\ and\ \bibinfo {author} {\bibfnamefont {A.}~\bibnamefont
  {Bhattacharjee}},\ }\href {\doibase 10.1063/1.873125} {\bibfield  {journal}
  {\bibinfo  {journal} {Physics of Plasmas}\ }\textbf {\bibinfo {volume} {5}},\
  \bibinfo {pages} {4028} (\bibinfo {year} {1998})}\BibitemShut {NoStop}%
\bibitem [{\citenamefont {Bogoyavlenskij}(2000)}]{Bogoyavlenskij2000}%
  \BibitemOpen
  \bibfield  {author} {\bibinfo {author} {\bibfnamefont {O.~I.}\ \bibnamefont
  {Bogoyavlenskij}},\ }\href {http://www.ncbi.nlm.nih.gov/pubmed/11017659}
  {\bibfield  {journal} {\bibinfo  {journal} {Physical Review Letters}\
  }\textbf {\bibinfo {volume} {84}},\ \bibinfo {pages} {1914} (\bibinfo {year}
  {2000})}\BibitemShut {NoStop}%
\bibitem [{\citenamefont {Low}(2006)}]{Low2006}%
  \BibitemOpen
  \bibfield  {author} {\bibinfo {author} {\bibfnamefont {B.~C.}\ \bibnamefont
  {Low}},\ }\href {\doibase 10.1086/506586} {\bibfield  {journal} {\bibinfo
  {journal} {The Astrophysical Journal}\ }\textbf {\bibinfo {volume} {649}},\
  \bibinfo {pages} {1064} (\bibinfo {year} {2006})}\BibitemShut {NoStop}%
\bibitem [{\citenamefont {Huang}\ \emph {et~al.}(2009)\citenamefont {Huang},
  \citenamefont {Bhattacharjee},\ and\ \citenamefont {Zweibel}}]{Huang2009}%
  \BibitemOpen
  \bibfield  {author} {\bibinfo {author} {\bibfnamefont {Y.-M.}\ \bibnamefont
  {Huang}}, \bibinfo {author} {\bibfnamefont {A.}~\bibnamefont
  {Bhattacharjee}}, \ and\ \bibinfo {author} {\bibfnamefont {E.~G.}\
  \bibnamefont {Zweibel}},\ }\href {\doibase 10.1088/0004-637X/699/2/L144}
  {\bibfield  {journal} {\bibinfo  {journal} {The Astrophysical Journal}\
  }\textbf {\bibinfo {volume} {699}},\ \bibinfo {pages} {L144} (\bibinfo {year}
  {2009})}\BibitemShut {NoStop}%
\bibitem [{\citenamefont {Janse}\ \emph {et~al.}(2010)\citenamefont {Janse},
  \citenamefont {Low},\ and\ \citenamefont {Parker}}]{Janse2010}%
  \BibitemOpen
  \bibfield  {author} {\bibinfo {author} {\bibfnamefont {A.~M.}\ \bibnamefont
  {Janse}}, \bibinfo {author} {\bibfnamefont {B.~C.}\ \bibnamefont {Low}}, \
  and\ \bibinfo {author} {\bibfnamefont {E.~N.}\ \bibnamefont {Parker}},\
  }\href {\doibase 10.1063/1.3474943} {\bibfield  {journal} {\bibinfo
  {journal} {Physics of Plasmas}\ }\textbf {\bibinfo {volume} {17}},\ \bibinfo
  {pages} {092901} (\bibinfo {year} {2010})}\BibitemShut {NoStop}%
\bibitem [{\citenamefont {Low}(2010)}]{Low2010}%
  \BibitemOpen
  \bibfield  {author} {\bibinfo {author} {\bibfnamefont {B.~C.}\ \bibnamefont
  {Low}},\ }\href {\doibase 10.1088/0004-637X/718/2/717} {\bibfield  {journal}
  {\bibinfo  {journal} {The Astrophysical Journal}\ }\textbf {\bibinfo {volume}
  {718}},\ \bibinfo {pages} {717} (\bibinfo {year} {2010})}\BibitemShut
  {NoStop}%
\bibitem [{\citenamefont {Longbottom}\ \emph {et~al.}(1998)\citenamefont
  {Longbottom}, \citenamefont {Rickard}, \citenamefont {Craig},\ and\
  \citenamefont {Sneyd}}]{Longbottom1998}%
  \BibitemOpen
  \bibfield  {author} {\bibinfo {author} {\bibfnamefont {A.~W.}\ \bibnamefont
  {Longbottom}}, \bibinfo {author} {\bibfnamefont {G.~J.}\ \bibnamefont
  {Rickard}}, \bibinfo {author} {\bibfnamefont {I.~J.~D.}\ \bibnamefont
  {Craig}}, \ and\ \bibinfo {author} {\bibfnamefont {A.~D.}\ \bibnamefont
  {Sneyd}},\ }\href {\doibase 10.1086/305694} {\bibfield  {journal} {\bibinfo
  {journal} {The Astrophysical Journal}\ }\textbf {\bibinfo {volume} {500}},\
  \bibinfo {pages} {471} (\bibinfo {year} {1998})},\ \Eprint
  {http://arxiv.org/abs/9903288} {arXiv:9903288 [astro-ph]} \BibitemShut
  {NoStop}%
\bibitem [{\citenamefont {Craig}\ and\ \citenamefont
  {Sneyd}(2005)}]{Craig2005}%
  \BibitemOpen
  \bibfield  {author} {\bibinfo {author} {\bibfnamefont {I.~J.~D.}\
  \bibnamefont {Craig}}\ and\ \bibinfo {author} {\bibfnamefont {A.~D.}\
  \bibnamefont {Sneyd}},\ }\href {\doibase 10.1007/s11207-005-1582-8}
  {\bibfield  {journal} {\bibinfo  {journal} {Solar Physics}\ }\textbf
  {\bibinfo {volume} {232}},\ \bibinfo {pages} {41} (\bibinfo {year}
  {2005})}\BibitemShut {NoStop}%
\bibitem [{\citenamefont {Pontin}\ and\ \citenamefont
  {Craig}(2005)}]{Pontin2005}%
  \BibitemOpen
  \bibfield  {author} {\bibinfo {author} {\bibfnamefont {D.~I.}\ \bibnamefont
  {Pontin}}\ and\ \bibinfo {author} {\bibfnamefont {I.~J.~D.}\ \bibnamefont
  {Craig}},\ }\href {\doibase 10.1063/1.1987379} {\bibfield  {journal}
  {\bibinfo  {journal} {Physics of Plasmas}\ }\textbf {\bibinfo {volume}
  {12}},\ \bibinfo {pages} {072112} (\bibinfo {year} {2005})}\BibitemShut
  {NoStop}%
\bibitem [{\citenamefont {Craig}\ and\ \citenamefont {Pontin}(2014)}]{Craig}%
  \BibitemOpen
  \bibfield  {author} {\bibinfo {author} {\bibfnamefont {I.}~\bibnamefont
  {Craig}}\ and\ \bibinfo {author} {\bibfnamefont {D.}~\bibnamefont {Pontin}},\
  }\href@noop {} {\bibfield  {journal} {\bibinfo  {journal} {The Astrophysical
  Journal}\ }\textbf {\bibinfo {volume} {788}},\ \bibinfo {pages} {177}
  (\bibinfo {year} {2014})}\BibitemShut {NoStop}%
\bibitem [{\citenamefont {Craig}\ and\ \citenamefont
  {Effenberger}(2014)}]{Craig2014}%
  \BibitemOpen
  \bibfield  {author} {\bibinfo {author} {\bibfnamefont {I.~J.~D.}\
  \bibnamefont {Craig}}\ and\ \bibinfo {author} {\bibfnamefont
  {F.}~\bibnamefont {Effenberger}},\ }\href {\doibase
  10.1088/0004-637X/795/2/129} {\bibfield  {journal} {\bibinfo  {journal} {The
  Astrophysical Journal}\ }\textbf {\bibinfo {volume} {795}},\ \bibinfo {pages}
  {129} (\bibinfo {year} {2014})}\BibitemShut {NoStop}%
\bibitem [{\citenamefont {Craig}\ and\ \citenamefont
  {Sneyd}(1986)}]{Craig1986}%
  \BibitemOpen
  \bibfield  {author} {\bibinfo {author} {\bibfnamefont {I.~J.}\ \bibnamefont
  {Craig}}\ and\ \bibinfo {author} {\bibfnamefont {A.~D.}\ \bibnamefont
  {Sneyd}},\ }\href@noop {} {\bibfield  {journal} {\bibinfo  {journal} {The
  Astrophysical Journal}\ }\textbf {\bibinfo {volume} {311}},\ \bibinfo {pages}
  {451} (\bibinfo {year} {1986})}\BibitemShut {NoStop}%
\bibitem [{\citenamefont {Craig}\ and\ \citenamefont
  {Litvinenko}(2005)}]{Craig2005b}%
  \BibitemOpen
  \bibfield  {author} {\bibinfo {author} {\bibfnamefont {I.~J.~D.}\
  \bibnamefont {Craig}}\ and\ \bibinfo {author} {\bibfnamefont {Y.~E.}\
  \bibnamefont {Litvinenko}},\ }\href {\doibase 10.1063/1.1854154} {\bibfield
  {journal} {\bibinfo  {journal} {Physics of Plasmas}\ }\textbf {\bibinfo
  {volume} {12}},\ \bibinfo {pages} {032301} (\bibinfo {year}
  {2005})}\BibitemShut {NoStop}%
\bibitem [{\citenamefont {Pontin}\ \emph {et~al.}(2009)\citenamefont {Pontin},
  \citenamefont {Hornig}, \citenamefont {Wilmot-Smith},\ and\ \citenamefont
  {Craig}}]{Pontin2009}%
  \BibitemOpen
  \bibfield  {author} {\bibinfo {author} {\bibfnamefont {D.~I.}\ \bibnamefont
  {Pontin}}, \bibinfo {author} {\bibfnamefont {G.}~\bibnamefont {Hornig}},
  \bibinfo {author} {\bibfnamefont {A.~L.}\ \bibnamefont {Wilmot-Smith}}, \
  and\ \bibinfo {author} {\bibfnamefont {I.~J.~D.}\ \bibnamefont {Craig}},\
  }\href {\doibase 10.1088/0004-637X/700/2/1449} {\bibfield  {journal}
  {\bibinfo  {journal} {The Astrophysical Journal}\ }\textbf {\bibinfo {volume}
  {700}},\ \bibinfo {pages} {1449} (\bibinfo {year} {2009})}\BibitemShut
  {NoStop}%
\bibitem [{\citenamefont {Candelaresi}\ \emph {et~al.}(2014)\citenamefont
  {Candelaresi}, \citenamefont {Pontin},\ and\ \citenamefont
  {Hornig}}]{Candelaresi2014}%
  \BibitemOpen
  \bibfield  {author} {\bibinfo {author} {\bibfnamefont {S.}~\bibnamefont
  {Candelaresi}}, \bibinfo {author} {\bibfnamefont {D.}~\bibnamefont {Pontin}},
  \ and\ \bibinfo {author} {\bibfnamefont {G.}~\bibnamefont {Hornig}},\ }\href
  {\doibase 10.1137/140967404} {\bibfield  {journal} {\bibinfo  {journal} {SIAM
  Journal on Scientific Computing}\ }\textbf {\bibinfo {volume} {36}},\
  \bibinfo {pages} {15} (\bibinfo {year} {2014})}\BibitemShut {NoStop}%
\bibitem [{\citenamefont {Zhou}\ \emph {et~al.}(2014)\citenamefont {Zhou},
  \citenamefont {Qin}, \citenamefont {Burby},\ and\ \citenamefont
  {Bhattacharjee}}]{Zhou2014}%
  \BibitemOpen
  \bibfield  {author} {\bibinfo {author} {\bibfnamefont {Y.}~\bibnamefont
  {Zhou}}, \bibinfo {author} {\bibfnamefont {H.}~\bibnamefont {Qin}}, \bibinfo
  {author} {\bibfnamefont {J.~W.}\ \bibnamefont {Burby}}, \ and\ \bibinfo
  {author} {\bibfnamefont {A.}~\bibnamefont {Bhattacharjee}},\ }\href {\doibase
  10.1063/1.4897372} {\bibfield  {journal} {\bibinfo  {journal} {Physics of
  Plasmas}\ }\textbf {\bibinfo {volume} {21}},\ \bibinfo {pages} {102109}
  (\bibinfo {year} {2014})}\BibitemShut {NoStop}%
\bibitem [{\citenamefont {Desbrun}\ \emph {et~al.}(2005)\citenamefont
  {Desbrun}, \citenamefont {Hirani}, \citenamefont {Leok},\ and\ \citenamefont
  {Marsden}}]{Desbrun}%
  \BibitemOpen
  \bibfield  {author} {\bibinfo {author} {\bibfnamefont {M.}~\bibnamefont
  {Desbrun}}, \bibinfo {author} {\bibfnamefont {A.~N.}\ \bibnamefont {Hirani}},
  \bibinfo {author} {\bibfnamefont {M.}~\bibnamefont {Leok}}, \ and\ \bibinfo
  {author} {\bibfnamefont {J.~E.}\ \bibnamefont {Marsden}},\ }\href@noop {}
  {\bibfield  {journal} {\bibinfo  {journal} {arXiv preprint math/0508341}\ }
  (\bibinfo {year} {2005})}\BibitemShut {NoStop}%
\bibitem [{\citenamefont {Newcomb}(1962)}]{Newcomb}%
  \BibitemOpen
  \bibfield  {author} {\bibinfo {author} {\bibfnamefont {W.~A.}\ \bibnamefont
  {Newcomb}},\ }\href@noop {} {\bibfield  {journal} {\bibinfo  {journal}
  {Nuclear Fusion Supplement}\ }\textbf {\bibinfo {volume} {2}},\ \bibinfo
  {pages} {451} (\bibinfo {year} {1962})}\BibitemShut {NoStop}%
\bibitem [{\citenamefont {Finn}\ and\ \citenamefont {Kaw}(1977)}]{Finn1977}%
  \BibitemOpen
  \bibfield  {author} {\bibinfo {author} {\bibfnamefont {J.~M.}\ \bibnamefont
  {Finn}}\ and\ \bibinfo {author} {\bibfnamefont {P.~K.}\ \bibnamefont {Kaw}},\
  }\href {\doibase 10.1063/1.861709} {\bibfield  {journal} {\bibinfo  {journal}
  {Physics of Fluids}\ }\textbf {\bibinfo {volume} {20}},\ \bibinfo {pages}
  {72} (\bibinfo {year} {1977})}\BibitemShut {NoStop}%
\bibitem [{\citenamefont {Longcope}\ and\ \citenamefont
  {Strauss}(1993)}]{Longcope}%
  \BibitemOpen
  \bibfield  {author} {\bibinfo {author} {\bibfnamefont {D.}~\bibnamefont
  {Longcope}}\ and\ \bibinfo {author} {\bibfnamefont {H.}~\bibnamefont
  {Strauss}},\ }\href@noop {} {\bibfield  {journal} {\bibinfo  {journal}
  {Physics of Fluids B: Plasma Physics}\ }\textbf {\bibinfo {volume} {5}},\
  \bibinfo {pages} {2858} (\bibinfo {year} {1993})}\BibitemShut {NoStop}%
\end{thebibliography}%

\end{document}